\documentstyle[aps,prl,multicol,epsfig]{revtex} 
\begin{document}
\title{ Random magnetic field and quasi-particle
    transports in the mixed state of high $ T_{c} $ cuprates }
\author{Jinwu Ye }
\address{ Department of Physics, University of Houston, Houston, TX, 77204 }
\date{\today}
\maketitle
\begin{abstract}
   By a singular gauge transformation, the quasi-particle
  transport in the mixed state of high $ T_{c} $
  cuprates is mapped into charge-neutral composite
  Dirac fermion moving in {\em short-range}
  correlated random scalar and {\em long-range } correlated
  vector potential. A fully quantum mechanical approach to
  longitudinal and transverse thermal conductivities 
  is presented. The semi-classical
  Volovik effect is presented in a quantum mechanical way.
  The quasi-particle scattering from the random magnetic field which was
  completely missed in all the previous semi-classical
  approaches is the dominant scattering mechanism at sufficient
  high magnetic field. The implications for experiments are discussed.
\end{abstract}
\begin{multicols}{2}
  The general problem of quasi-particle transport in the mixed state of
  high $ T_{c} $ cuprates 
  is important, because simultaneous measurements of thermal 
  conductivities  $\kappa_{xx} $ and $\kappa_{xy} $
 provide a lot of information on the new physics pertinent to
  $ d $ wave superconductors.  On the experimental side,
 Krishana {\sl et al} observed that in superconducting BSCCO and YBCO,
 at temperature $ T > 5K $, the longitudinal thermal conductivity 
 $ \kappa_{xx}(H) $ initially decreases with applied magnetic field $ H $,
 then reaches a plateau \cite{ong1}. They also measured \cite{ong2,new}
 thermal Hall conductivity $ \kappa_{xy} $ at $ T > 10K $  and extracted
 the thermal Hall angle $ \tan \theta=\frac{\kappa_{xy}}{\kappa_{xx}} $.
 On the theoretical
 side, employing semi-classical approximation, Volovik pointed out that  
 the circulating supercurrents around vortices
 induce Doppler energy shift to the quasi-particle spectrum, which leads to
 a finite density of states at the nodes \cite{volovik}. 
 This effect ( Volovik effect) has been employed
 to explain the above experimental observations of $ \kappa_{xx} $
 by several authors \cite{hirs}. However, semi-classical method can
 not be used to calculate $ \kappa_{xy} $. A fully quantum mechanical
 approach is needed to get any information on $ \kappa_{xy} $.
 
 Starting from BCS Hamiltonian, Wang and MacDonald performed
 a first numerical calculation on quasi-particle spectrum in vortex lattice
 state \cite{mac}.  By phenomenological scaling argument,
 Simon and Lee (SL)\cite{lee} proposed 
 the approximate scaling forms for $ \kappa_{xx} $ and $ \kappa_{xy} $
 for {\em dirty} $ d $ wave superconductors in the mixed state.
 Anderson \cite{and} employed a singular gauge transformation to study
 quasi-particle dynamics in the mixed state.
 Franz and Tesanovic (FT) employed a different singular gauge transformation
 to map the quasi-particle in a square vortex lattice state to Dirac fermion
 moving in an effective periodic scalar and vector potential with {\em zero}
 average and studied the quasi-particle spectrum \cite{sing}.
  Using FT
 transformation, Marinelli {\sl et al} studied the spectrum in various
 kinds of vortex lattice states in great detail \cite{bert1}.

  In this paper, by considering carefully the gauge invariance overlooked
  by previous authors \cite{lee,sing}, we extend FT singular gauge
  transformation to include the curvature term which is important to
  $ \kappa_{xy} $. After clarifying
  some important subtle points of the singular gauge transformation, we
  prove exactly that there is {\em no} Landau level quantitation, then
  apply it to {\em disordered} vortex state with logarithmic interaction
  between vortices. We find
  that quasi-particle moving in the disordered
  vortex state of high $ T_{c} $ cuprates can be mapped into charge-neutral
  composite Dirac fermion moving in {\em short-range}
  correlated random scalar potential and {\em long-range} correlated
  vector potential with zero average.
 The quasi-particle scattering
  from the long-range correlated internal gauge field dominates over
  those from the well-known Volovik effect and the non-magnetic scattering
  at sufficient high magnetic field $ H \gg H^{*}\sim
  H_{c2} (\frac{T}{T_{c}})^{2} $.
  $ \kappa_{xx} $ satisfies scaling Eq.\ref{xx}, at $ H > H^{*} $,
  it approaches  {\em different} pinning strength dependent plateaus
  due to the vertex correction shown in Fig.1.
  However, $ \kappa_{xy} $ satisfies scaling Eq.\ref{xy}, at $ H> H^{*} $, it
  increases as $ \sqrt{H} $.

  We start from the $ d $ wave BCS Hamiltonian in the presence of external
  magnetic field:
\begin{equation}
      H=\int d x
      d^{\dagger}(x) 
     \left ( \begin{array}{cc}
		h + V(x)  &   \hat{\Delta}   \\
		\hat{\Delta}^{\dagger}   &   -h^{*} -V(x)  \\
		\end{array}   \right )  d(x)
\label{start}
\end{equation}
   with $ d_{\uparrow}(x)= c_{\uparrow}(x), d_{\downarrow}(x)
     = c^{\dagger}_{\downarrow}(x) $,
   $ V(x) $ is random chemical potential due to {\em non-magnetic} impurities.

   The non-magnetic impurity scattering part can be written as:
\begin{equation}
   H_{imp}= V(x) d^{\dagger}_{\alpha}(x) \tau^{3}_{\alpha \beta} d_{\alpha}(x)
\label{imp}
\end{equation}

     We assume that $ V(x) $ satisfies Gaussian distribution with
     zero mean and variances $ \Delta_{A} $
\begin{equation}
   < V(x) >=0,~~~~ <V(x) V(x^{\prime}) >= \Delta_{A} \delta^{2}(x-x^{\prime})
\label{zerof}
\end{equation}

     The gauge invariance dictates:
\begin{eqnarray}
  h   & = & \frac{1}{2m} (\vec{p}-\frac{e}{c} \vec{A} )^{2} -\epsilon_{F}   \nonumber    \\
  \hat{\Delta} &=& \frac{1}{4 p_{F}^{2}}[
  \{p_{x}-\frac{1}{2} \partial_{x}\phi, p_{y}-\frac{1}{2} \partial_{y} \phi\}
     \Delta(\vec{r})
             \nonumber  \\
  &+& (p_{x}-\frac{1}{2} \partial_{x} \phi) \Delta(\vec{r}) 
  (p_{y}+\frac{1}{2} \partial_{y} \phi)
         \nonumber  \\
  &+&(p_{y}-\frac{1}{2} \partial_{y} \phi) \Delta(\vec{r}) 
   (p_{x}+\frac{1}{2}\partial_{x} \phi)
                    \nonumber  \\
   & + & \Delta(\vec{r}) \{p_{x}+\frac{1}{2} \partial_{x} \phi, 
   p_{y}+\frac{1}{2}\partial_{y} \phi \} ] 
\label{gauge}
\end{eqnarray}
  where $\phi $ is the phase of $ \Delta(\vec{r}) $ \cite{mistake}.

      The gauge invariance in $ \hat{\Delta} $ 
   has not been taken into account in Refs.\cite{lee,sing}. 
   Although its correct treatment does not affect
     the linearized Hamiltonian ( see Eq.\ref{linear} ), it is crucial
    to the curvature term ( see Eq.\ref{curv} ) which is important
     to $ \kappa_{xy} $. The similar gauge invariance was considered
     in Eq.2.12 of Ref.\cite{jinwu}.


  Following FT \cite{sing}, we introduce {\em composite fermion } \cite{bert2} 
 $ d_{c} $ by performing a singular unitary transformation $ d=Ud_{c} $:
\begin{equation}
    H_{s}=U^{-1} H U,~~~~~~~~U=
     \left ( \begin{array}{cc}
		e^{i \phi_{A}(\vec{r})}  &   0  \\
		0  &   e^{-i \phi_{B}(\vec{r}) }  \\
		\end{array}   \right )
\end{equation}
   where $ \phi_{A} $ is the phase from the vortices in sublattice $ A $
   and $ \phi_{B} $ is the phase from the vortices in sublattice $ B $.

   In this letter, we assume the thermal currents are sufficiently weak
  that the vortices remain
  pinned by non-magnetic impurities. Therefore, the transport properties 
  of $ d $
  is  exactly the same as the composite fermions $ d_{c} $.
   It is easy to check that $ U $ {\em commutes} with $ \tau^{3} $, therefore
   the transformation leaves Eq.\ref{imp} $ H_{imp} $ {\em invariant}.

     Expanding $ H_{s} $ around the node 1 where
     $ \vec{p}=(p_{F},0) $,
     we obtain $ H_{s}
     =H_{l}+H_{c} $ where the linearized Hamiltonian
     $ H_{l} $ is given by:
\begin{eqnarray}
   H_{l} &= & v_{f} (p_{x}+  a_{x} ) \tau^{3} 
   +v_{2} (p_{y}+  a_{y} ) \tau^{1} 
   + v_{f}  v_{x}(\vec{r})
                  \nonumber \\
    & + & V(x) \tau^{3} + (1 \rightarrow 2, x \rightarrow y )
\label{linear}
\end{eqnarray}
   where $ \vec{v}_{s}= \frac{1}{2} (\vec{v}^{A}_{s}+ \vec{v}^{B}_{s})
   = \frac{\hbar}{2} \nabla \phi-\frac{e}{c} \vec{A}$ is the total
     superfluid {\em momentum} and
 $ a_{\alpha}= \frac{1}{2}( v^{A}_{\alpha}-v^{B}_{\alpha})
 = \frac{1}{2}( \nabla \phi_{A}-\nabla \phi_{B}) $ is the
   internal gauge field. Anderson's gauge choice
   is $ \phi_{A}=\phi, \phi_{B}=0 $ or vice versa \cite{and,single}.

    We get the corresponding expression at node $ \bar{1} $ and $ \bar{2} $
  by changing $ v_{f} \rightarrow -v_{f}, v_{2} \rightarrow -v_{2} $
  in the above Eq.

    The curvature term $ H_{c} $ can be written as:
\begin{equation}
   H_{c} =  \frac{1}{m} [  \{ \Pi_{\alpha}, v_{\alpha} \}
	  +  \frac{  \vec{\Pi}^{2}+ \vec{v}^{2} }{2} \tau^{3} 
	  + \frac{\Delta_{0}}{2 \epsilon_{F}} \{ \Pi_{x}, \Pi_{y}\} \tau^{1} ]
\label{curv}
\end{equation}
     Where $ \vec{\Pi}= \vec{p}+  \vec{a} $ is the covariant
     derivative. $ H_{c} $ takes the {\em same } form for all the four nodes.

   It is easy to see that $ v_{\alpha} (\vec{r}) $ acts as a scalar
   scattering potential, it respects time-reversal (T) symmetry, but
   breaks Particle-Hole (PH) symmetry \cite{lud}.
   There are two {\em very different} kinds of internal
   gauge fields in $ H_{l} $:
   $ V(x) $ is due to non-magnetic impurity scattering at zero field,
   $ a_{\alpha} $ is completely due to Aharonov and Bohm (AB)
   phase scattering \cite{single}
   from vortices generated by external magnetic field.
   They both respect P-H symmetry. In general, $ V(x) $ breaks T symmetry,
   but T symmetry is restored in the unitary limit \cite{pepin}.
   For general flux quantum $ \alpha $, $ a_{\alpha} $ breaks T symmetry, 
   but T symmetry is restored at $ \alpha=1/2 $ \cite{thin}
   because $ \alpha=1/2 $ flux
   quantum is equivalent to $ \alpha=-1/2 $ one due to the
   periodicity under $ \alpha \rightarrow \alpha+1 $. This is also
   the underlying physical reason why we are able to choose the two
   sublattices $ A $ and $ B $ freely without changing any physics.
   Due to this exact T symmetry, there is no Landau level quantization
   as claimed in \cite{and}.

   In $ H_{c} $, the only term which breaks both P-H and T symmetry \cite{lud}
   is  $ \psi^{\dagger} \{p_{\alpha}, v_{\alpha} \} \psi
      = -i v_{\alpha} ( \nabla_{\alpha} \psi^{\dagger} \psi-
        \psi^{\dagger} \nabla_{\alpha} \psi ) $. This term will lead to
   thermal Hall conductivity to be discussed in the following \cite{mass}.

    From Eq.\ref{linear}, it is easy to identify the conserved
    charge currents at node 1:
\begin{equation}
   j_{1x} = \psi^{\dagger}_{1}(x)  v_{F} \tau^{3} \psi_{1}(x)
  ,~~~~~~j_{1y} = \psi^{\dagger}_{1}(x)  v_{2} \tau^{1} \psi_{1}(x)
\label{gap}
\end{equation}
   with the currents at node 2 differing from the above expressions by 
   $ (1 \rightarrow 2, x \rightarrow y ) $.
    It is known that the {\em  charge} conductivity of $ d_{c} $
    corresponds to  the {\em spin} conductivity of $ c $ electrons 
     Because the spin $ \sigma_{s} $ and thermal conductivities
    are related by Wiedemann-Franz law \cite{nice,senthil}.
    For simplicity, in the following, we only give specific expressions for
    the spin conductivity which can be evaluated by Kubo formula.

   The Hamiltonian $ H_{l} + H_{c} $ enjoys
   gauge symmetry $ U_{u}(1) \times U_{s}(1) $,
   the first being uniform (or external ) and the second being 
   staggered ( or internal ) gauge symmetry.
      Although the composite fermion $ d_{\alpha} $
      is charge {\em neutral} to the external magnetic field, it carries
      charge $ 1 $ to the internal gauge field $ a_{\alpha} $.

  We assume a randomly pinned vortex array with logarithmic interaction
  between vortices. In the {\em hydrodynamic limit}, after averaging over
  all the possible positions of the vortices $ { R_{i} } $,
  we find:
\begin{eqnarray}
   & &  <v_{\alpha}> =  < a_{\alpha} >=0,~~~~~
      <v_{\alpha}(\vec{k}) a_{\beta}( \vec{k}^{\prime} ) >=0
                          \nonumber   \\
   & &<v_{\alpha}(\vec{k}) v_{\beta}( -\vec{k} ) >
     = \pi^{2} ( \delta_{\alpha \beta} - \frac{ k_{\alpha} k_{\beta}}
     {k^{2} }) \frac{n_{v}}{ k^{2}+ n_{v} }   
            \nonumber  \\
  & & <a_{\alpha}(\vec{k}) a_{\beta}( - \vec{k} ) >
     = \pi^{2} ( \delta_{\alpha \beta} - \frac{k_{\alpha} k_{\beta}}
   { k^{2} } ) \frac{n_{v}}{ k^{2} }   
\label{second}
\end{eqnarray}
     Where the vortex density is $ n_{v}= \frac{N}{V}= \frac{H}{H_{c2}} 
     \frac{1}{ \xi^{2} } $.

   The first line in Eq.\ref{second} is exact. The $ v-v $ and
   $ a-a $ correlators are the most general forms consistent with
   the incompressibility of the vortex system \cite{andy}. The pinning
   strength will only enter as prefactors in front of $ n_{v} $.
   For notational simplicity, we suppress these prefactors.  

   Because $ v $ and $ a $ are decoupled at quadratic order (see Eq.9),
   the long-range logarithmic interaction
   between vortices suppresses the fluctuation of superfluid velocity,
   but does {\em not} affect the fluctuation of the internal gauge field.
   Therefore the scalar field acquires a " mass " determined by the vortex
   density, but the gauge field remains  " massless ".
    The gauge field is a pure quantum mechanical effect, it was completely
   missed in all the previous semi-classical approaches 
   \cite{volovik,hirs}.
   Here we explicitly demonstrate that being gapless,
   its fluctuation even {\em  dominates } over the well-known Volovik effect
   in the low energy limit.
    It also dominates over that from the non-magnetic
    scattering at sufficiently high field and low energy limit.

  In fact, in the weakly type II limit $ \xi < \lambda < d_{v} $,
  the superfluid velocity vanishes in the interior of superconductor,
  the long-range
  correlated gauge potential in Eq.\ref{second} becomes the only scattering
  mechanism \cite{weak}.

     From the Eq.\ref{second}, it is easy to realize that the internode
     scattering $\sim k^{-2}_{F} $ is weaker than the intra-node
     scattering $ \sim p^{-2}_{0}  $ by a factor $ \alpha^{-1}
     = \frac{\Delta_{0}}{\epsilon_{F}} \ll 1 $. In the following
     we neglect the internode scattering.

      Up to the order of Gaussian cumulants, the scalar potential
      and vector potential are uncorrelated. However, they are
      correlated by Non-Gaussian cumulants. The lowest order non-Gaussian
      cumulants are the skewness: 
\begin{eqnarray}
 & & <v_{\alpha}(\vec{r}_{1}) v_{\beta}(\vec{r}_{2} ) a_{\gamma}( \vec{r}_{3} ) >
   = <a_{\alpha}(\vec{r}_{1}) a_{\beta}(\vec{r}_{2} ) a_{\gamma}( \vec{r}_{3} ) >
   =0               \nonumber   \\
  & & <v_{\alpha}(\vec{k}_{1}) v_{\beta}(\vec{k}_{2} ) 
   v_{\gamma}( \vec{k}_{3} ) >
     =  \pi^{3} n_{v} \delta(\vec{k}_{1} +\vec{k}_{2} +\vec{k}_{3} )
               \nonumber  \\
   & & \times  \frac{ -i \epsilon_{\alpha \delta} k_{1 \delta}(
      \vec{k}_{2} \cdot \vec{k}_{3} \delta_{\beta \gamma} - 
      k_{2 \gamma} k_{3 \beta} ) } 
      { ( k^{2}_{1}+ n_{v} ) ( k^{2}_{2}+ n_{v} )
       ( k^{2}_{3}+ n_{v} ) }
                 \nonumber   \\
  & & <v_{\alpha}(\vec{k}_{1}) a_{\beta}(\vec{k}_{2} ) 
   a_{\gamma}( \vec{k}_{3} ) > 
     =  \pi^{3} n_{v} \delta(\vec{k}_{1} +\vec{k}_{2} +\vec{k}_{3} )
      \nonumber   \\
   & & \times \frac{ -i \epsilon_{\alpha \delta} k_{1 \delta}(
      \vec{k}_{2} \cdot \vec{k}_{3} \delta_{\beta \gamma} - 
      k_{2 \gamma} k_{3 \beta} ) } 
      { ( k^{2}_{1}+ n_{v} )  k^{2}_{2}  k^{2}_{3} } 
\label{third}
\end{eqnarray}

  In fact, because {\em any} distribution function satisfies
 $ P[ a_{\alpha}(x) ]=P[ -a_{\alpha}(x) ] $, {\sl any correlators 
 involving odd number of $ a_{\alpha} $ vanish}. After coarse graining,
 the exact T symmetry of $ H_{l} $ is approximated by the average one.
 This approximation will lead to correct behaviors of
 self-averaging physical quantities. But it does not apply to
 non-self-averaging quantity such as Hall conductance fluctuation. 
 
\underline{ The discussion on $ \kappa_{xx} $: }
    The scalar potential capture the essential physics of Volovik effect:
    the quasi-particles energies are shifted by superfluid flow.
    Following the RG analysis in Ref.\cite{lud}, it can be shown
     that the random scalar potential is
    {\em marginally relevant}, therefore generates finite density of states
    at {\em zero} energy. Because $ k_{F} l_{tr} \sim k_{F} d_{v}
    \gg k_{F} \xi \sim 5 $, the SCBA in standard impurity scattering
      process can be applied to calculate the low energy scattering rate.
     In $ k, \omega \rightarrow 0  $ limit, it leads to
\begin{equation}
 1= v^{2}_{F}  \pi^{2} n_{v} \int \frac{d^{2} p}{ (2 \pi)^{2}}
       \frac{ p^{2}_{y} }
       { p^{2}( p^{2}+ n_{v}) } \frac{1}{ \Gamma^{2}_{0} + E^{2}_{p}}
\label{self}
\end{equation}
      where $ E^{2}_{k}= (v_{f} k_{1} )^{2} +(v_{2} k_{2})^{2} $,
     the retarded self-energy $ \Sigma^{R}(k, \omega)= 
 \Sigma(\vec{k}, i \omega \rightarrow \omega + i \delta ) $,
     the zero energy scattering rate is $ \Gamma_{0}= -Im\Sigma^{R}(0,0) $.

     The above equation leads to the quasi-particle lifetime:
\begin{equation}
   1/\tau_{l} \sim \Gamma_{0} \sim  \Delta_{0} \sqrt{ \frac{H}{H_{c2}}} 
\label{cpa}
\end{equation}

   Now we look at the vertex correction to $ \kappa_{xx} $ due to the 
   ladder diagram shown in Fig.1 at $ T=0 $.

\vspace{0.25cm}

\epsfig{file=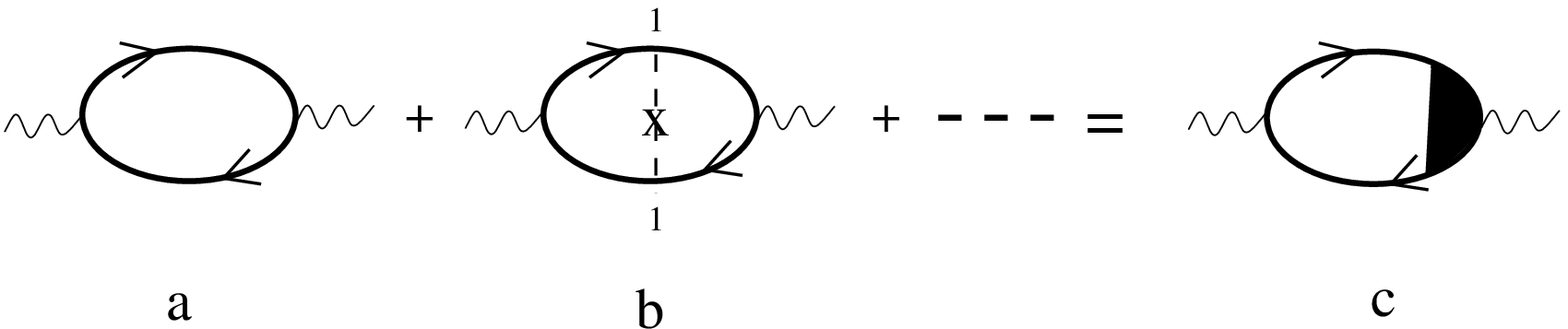,width=3.2in,height=0.8in,angle=0}

{\footnotesize {\bf Fig 1:} Vertex correction to longitudinal conductivity}    

\vspace{0.25cm}

   Fig.1a is just the bubble diagram. It leads
   to the well known bubble conductivity \cite{nice,senthil}:
 $\sigma^{0}_{xx}= \frac{s^{2}}{\pi^{2}}
  \frac{ v^{2}_{f} + v^{2}_{2} }{v_{f} v_{2} } $.

   By checking the integral equation satisfied by the vertex function
   $ \Gamma(p, i\omega, i\omega+i \Omega) = \tau^{3} (1+
  \Lambda(p, i\omega, i\omega+i \Omega)) $
  depicted in Fig.1c,  we find $ \Lambda $ is at the order of 1
  independent of $ n_{v} $, therefore $ \sigma_{xx} $ receives vertex
  correction of order 1 independent of $ n_{v} $ \cite{num}.
  This vertex correction was completely missed in the semiclassical treatments
  \cite{hirs}.

 The random gauge field gives {\em additional} scattering mechanism.
 it has scaling dimension 2, therefore strongly relevant \cite{nick} and
 {\em dominates} over Volovik effect at low
 energy limit. Similar SCBA to Eq.\ref{self} leads to {\em
 logarithmic divergent } quasi-particle scattering rate
 $ 1/\tau_{l} $ which is not gauge-invariant anyway.
  But the vertex correction similar to Fig.1 removes
  the logarithmic divergence and leads to
  the finite gauge-invariant transport time $ \tau_{tr} 
 \sim \sqrt{\frac{H_{c2}}{H}} \frac{1}{\Delta_{0}} $ \cite{num}.

     The vertex correction to the bubble conductivity 
 due to the non-magnetic impurity scattering Eq.\ref{zerof} among the four nodes
 was calculated in Ref.\cite{nice}, it was found to be negligible.
 It is obvious that the two vertex corrections are different due to
 different scattering mechanisms, therefore the two conductivity
 values are {\em different}, although the bubble results are the same.

   At finite temperature, the $ T $ dependence comes solely from
   the Fermi function, $ \sigma_{xx} $ should satisfy the
   following scaling ( $ T_{c} \sim \Delta_{0} $ ):
\begin{equation}
   \sigma_{xx}(H, T )= F_{1}( a T \tau_{tr} )=
    F_{1}( a \frac{T}{T_{c}} \sqrt{\frac{H_{c2}}{H}} )
\label{xx}
\end{equation}

   This scaling is consistent with Simon and Lee \cite{lee} using
   phenomenological scaling argument. Our derivation bring out {\em explicitly}
   the underlying physical process: the quasiparticle scattering
   due to the long-range correlated random gauge potential.
  Pushing further, we conclude that 
  in the high field limit $ H \gg a^{2} H_{c2}
  (\frac{T}{T_{c}} )^{2} $, $ \sigma_{xx} $ should  approach
  the $ T=0 $ value $ F_{1}(0)= \sigma_{xx}(0) $ at the order of 1.
  This value depends on not only the anisotropy parameter
  $ \alpha= v_{f}/v_{2} $, but also the pinning strength appearing in
  Eq.\ref{second}. This dependence could explain the {\em different}
  plateau values observed in the experiments \cite{ong1}.

   Unfortunately, the value $ \sigma_{xx}(0) $ is hard to
  be sorted out experimently due to the large background contributions
  from phonons \cite{ong1,ong2}. 

\underline{ The discussion on $ \kappa_{xy} $: }
    We start with $ H_{l} $. In order to
  get a non-vanishing $ \kappa_{xy} $, we must identify terms which
  break {\em both} T and P-H symmetry. As shown previously, $ H_{l} $
  respects exact T symmetry, therefore $ < \sigma_{xy}> =0 $ to the
 linear order.
  We have to go to the curvature term Eq.\ref{curv} to see its contribution
  to $ <\sigma_{xy} > $.

   The first contribution comes from
  the skewness between Volovik term and the
   $ \{ p_{\alpha},v_{\alpha} \} $  term
   $ < \partial v_{\alpha}(\vec{r}_{1}) 
  v_{\beta}(\vec{r}_{2} ) v_{\gamma}( \vec{r}_{3} ) > $ in Eq.\ref{third}
  which breaks both T and P-H symmetry.

   Just like in $ \kappa_{xx} $, the skewness between scalar and
   random gauge field $ < \partial v_{\alpha}(\vec{r}_{1}) 
  a_{\beta}(\vec{r}_{2} ) a_{\gamma}( \vec{r}_{3} ) > $ in Eq.\ref{third}
 gives {\em additional} scattering mechanism. 
 It even {\em dominates} over the pure scalar skewness at low
 energy limit.

  Due to the antisymmetric tensor $ \epsilon_{\alpha \delta} $,
  we find $ \sigma_{2xy}=-\sigma_{1yx}= \sigma_{1xy} $.
  Because both skewnesses are even under $ v_{f} \rightarrow -v_{f},
  v_{2} \rightarrow v_{2} $, it is easy to find that $ \sigma_{1xy}=
  \sigma_{\bar{1}xy} $, therefore $ \sigma_{xy}
   =\sigma_{1xy}+\sigma_{2xy} +\sigma_{\bar{1}xy}+\sigma_{\bar{2}xy}
   = 4 \sigma_{1xy} $. 

  Because the $ \{ p_{\alpha},v_{\alpha} \} $ term contains one
   more derivative, simple power counting leads to
\begin{equation}
   <\sigma_{xy}(H, T )>= \frac{T_{c}}{\epsilon_{F}} \sqrt{\frac{H}{H_{c2}}}
      F_{2}( b \frac{T}{T_{c}} \sqrt{\frac{H_{c2}}{H}} )
\label{xy}
\end{equation}
   
   This scaling is consistent with Simon and Lee \cite{lee} by
    phenomenological scaling argument. Our derivation bring out {\em explicitly}
   the leading contributions from the $\{ p_{\alpha}, v_{\alpha} \} $
   term in the curvature term and also its
   small numerical factor $ \frac{T_{c}}{\epsilon_{F}} $.
  Pushing further, we conclude that in high field limit $ H \gg b^{2}
  H_{c2} (\frac{T}{T_{c}})^{2} $, $ \sigma_{xy} $ should increase with $ H $
  as $ \frac{T_{c}}{\epsilon_{F}} \sqrt{\frac{H}{H_{c2}}} F_{2}(0) $.
   Taking $ \alpha \sim 10, H \sim 10T, H_{c2} \sim 150T $, we find
   the prefactor is about $ 1/40 $, so $ \kappa_{xy} $ is smaller
  than $ \kappa_{xx} $
  by a factor of $1/40 $. The smallness of $ \kappa_{xy} $ make it
  difficult to measure experimentally.

   The most recent data at $ 10K < T < 30K $ for $ \kappa_{xy}/T^{2} $
   \cite{new} was shown to satisfy quite well the scaling $ \kappa_{xy}/T^{2}
   = F( b \sqrt{H}/T ) $ which follows from
   Eq.\ref{xy}, however, it continues to decrease up to
   $ H=14T $ instead of increasing linearly with $ \sqrt{H}/T $
     as follows from Eq.\ref{xy}. The discrepancy may be due to the 
    inelastic scattering at $ T > 10K $ not considered in this paper.
    For technical reason, so far the data is not available at  $ T< 10K $.

  In conclusion, we point out that the dominant scattering mechanism
  at sufficient high magnetic field is due to the long-range correlated
  random gauge potential instead of the well-known Volovik effect.

 This work was supported by NSF Grant No. DMR-97-07701 and
 university of Houston. I am deeply indebted to
  B. I. Halperin, A. J. Millis, N. Read and Z. Tesanovic for very
  valuable discussions. I thank D. Arovas, A. V. Balatsky,
  A. W. W. Ludwig, A. D. Stone and
  A. M. Tsvelik for patiently explaining their work to him. I also thank
  C. Chamon, M. Franz, T. Senthil and C. S. Ting for helpful discussions.

\end{multicols}

\begin{references}
\bibitem{ong1} K. Krishana {\sl et al}, Science 277, 83 (1997);
    H. Aubin {\sl et al} Science 280,9a (1998), Phys. Rev. Lett.
    82, 624 (1999).

\bibitem{ong2} K. Krishana {\sl et al}, Phys. Rev. Lett. 82, 5108 (1999).

\bibitem{new} N. P. Ong, private communication.

\bibitem{volovik} G. E. Volovik, Sov. Phys. JETP 58, 469 (1993)

\bibitem{hirs} C. Kubert and P. J. Hirschfeld, Phys. Rev. Lett. 80, 
 4963 (1998).  M. Franz, {\sl ibid} 82, 1760 (1999).

\bibitem{mac} Y. Wang and A. H. MacDonald, Phys. Rev. B52, R3876 (1995).

\bibitem{lee} S. H. Simon and P. A. Lee, Phys. Rev. Lett. 78, 1548 (1997)

\bibitem{and} P. W. Anderson, cond-mat/9812063.

\bibitem{sing} M. Franz and Z. Tesanovic, Phys. Rev. Lett. 84, 554 (2000);

\bibitem{bert1} Luca Marinelli, B. I. Halperin  and S. H. Simon, 
   Phys. Rev. B {\bf 62}, 3488 (2000).

\bibitem{nice}  A. C. Durst and P. A. Lee, 
   Phys. Rev. B {\bf 62}, 1270 (2000).

\bibitem{mistake} In the first version of the paper, the external gauge
  field $ \vec{A} $ was used to fix the gauge invariance of
  $ \hat{\Delta} $ operator. I thank Franz and Tesanovic for pointing out
  the mistake to me \cite{mass}.
  The lack of gauge invariance in the $\hat{\Delta} $ operator
  in Refs.\cite{lee,sing} was also independently realized and
  discussed by N. Read. See also O. Vafek {\sl et al}, cond-mat/0007296.


\bibitem{jinwu} Jinwu Ye and S. Sachdev, Phys. Rev. B44, 10173 (1991);

\bibitem{bert2} To some extent, the composite fermion here is similar
   to that introduced in FQHE by B.I.~Halperin, P.A.~Lee
   and N.~Read, Phys. Rev. B {\bf 47}, 7312 (1993).

\bibitem{single} In a single vortex limit, Anderson\cite{and}'s gauge
     has to be used, then $ a_{\alpha}=\frac{1}{2} \nabla \phi $ (or
    $ a_{\alpha}= -\frac{1}{2} \nabla \phi $ ) with $ \nabla \times \nabla
  \phi= 2 \pi \hat{z} \delta(\vec{r}) $. Eq.\ref{linear}
   describes the quasi-particle scattering from $\alpha=1/2 $ flux quantum,
   in addition to the superfluid shift ( Volovik effect ).
    The strong quasi-particle scattering ( close to the unitary limit)
    from a single magnetic string
    with general $ \alpha $ flux quantum was discussed in the pioneering
    paper by Y. Aharonov and D. Bohm, Phys. Rev. 115, 485 (1959);
   for a review, see S. Olariu and I. I. Popescu, Rev. Mod. Phys.
   57, 339 (1985).

\bibitem{weak} A. K. Geim, S. J. Bending and I. V. Grigorieva,
   Phys. Rev. Lett. 69, 2252 (1992).



\bibitem{lud} A. W. W. Ludwig {\sl et al},
  Phys. Rev. B {\bf 48}, 13749, (1993); A. A. Nersesyan, A. M. Tsvelik and
  F. Wenger, Phys. Rev. Lett. 72, 2628 (1994), Nucl. Phys. B438, 561 (1995);
   Jinwu Ye, Phys. Rev. B60, 8290 (1999).
  Note this P-H symmetry is within a single node which is different
  from the P-H symmetry coming from
  spin $ SU(2) $ symmetry which relates the two opposite nodes
  $ i $ and $ \bar{i} $.

\bibitem{pepin} C. Pepin and P. A. Lee, cond-mat/0002227.

\bibitem{thin}  In fact, this is only true when we neglect the finite
    core size $ \sim \xi $ of the vortex.

\bibitem{mass}  If $ \vec{A} $ were used in $ \hat{\Delta} $ operator
  in Eq.\ref{gauge},
  there would be an extra random mass term $ -\frac{\Delta_{0}}{8 \epsilon_{F} }
  (\partial_{x} v_{sy}+ \partial_{y} v_{sx} ) \tau^{2} $ inside the bracket in
  Eq.\ref{curv}.  The mass term also breaks both P-H and T, but is supressed
  by a factor $ \frac{\Delta_{0}}{8 \epsilon_{F}} $ compared to 
  $ \{p_{\alpha}, v_{\alpha} \} $, therefore can be {\em dropped out} anyway.



\bibitem{senthil} T. Senthil, M.P. A. Fisher, L. Balents and C. Nayak,
Phys. Rev. Lett. {\bf 81}, 4704 (1998).

\bibitem{andy} I thank A. J. Millis for convincing me that the length scale
    beyond which
   the vortices are incompressible is the mean distance between vortices $ d_{v}
    = \xi \sqrt{H_{c2}/H} $ instead of the penetration length $ \lambda $.

\bibitem{num} Jinwu Ye, unpublished.

\bibitem{nick} The randomly placed vortices in the context of paired FQHE
   was independently discussed by N. Read and D. Green, 
   Phys. Rev. B {\bf 61}, 10267 (2000).






\end{references}
\end{document}